\documentclass[utf8]{frontiersSCNS} 

\usepackage{url,lineno,microtype,subcaption}
\usepackage[onehalfspacing]{setspace}
\usepackage{booktabs} 

\newcommand{\hl}[1]{#1}

\def\keyFont{\fontsize{8}{11}\helveticabold }
\def\firstAuthorLast{Tran {et~al.}} 
\def\Authors{Tram Thi Minh Tran\,$^{*}$, Callum Parker\,, Yiyuan Wang\, and Martin Tomitsch\,}
\def\Address{Design Lab, Sydney School of Architecture, Design and Planning, The University of Sydney, Sydney, NSW 2008, Australia}
\def\corrAuthor{Tram Thi Minh Tran}
\def\corrEmail{tram.tran@sydney.edu.au}

\begin{document}
\onecolumn
\firstpage{1}



\title{Designing Wearable Augmented Reality Concepts to Support Scalability in Autonomous Vehicle–Pedestrian Interaction}

\author[\firstAuthorLast ]{\Authors} 
\address{} 
\correspondance{} 

\extraAuth{}

\maketitle

\begin{abstract}

Wearable augmented reality (AR) offers new ways for supporting the interaction between autonomous vehicles (AVs) and pedestrians due to its ability to integrate timely and contextually relevant data into the user’s field of view. This paper presents novel wearable AR concepts that assist crossing pedestrians in multi-vehicle scenarios where several AVs frequent the road from both directions. Three concepts with different communication approaches for signaling responses from multiple AVs to a crossing request, as well as a conventional pedestrian push button, were simulated and tested within a virtual reality environment. The results showed that wearable AR is a promising way to reduce crossing pedestrians’ cognitive load when the design offers both individual AV responses and a clear signal to cross. The willingness of pedestrians to adopt a wearable AR solution, however, is subject to different factors, including costs, data privacy, technical defects, liability risks, maintenance duties, and form factors. We further found that all participants favored sending a crossing request to AVs rather than waiting for the vehicles to detect their intentions – pointing to an important gap and opportunity in the current AV–pedestrian interaction literature. 

\tiny
 \keyFont{ \section{Keywords:} autonomous vehicles, vehicle-to-pedestrian communication, external human-machine interfaces, user-initiated communication, vulnerable road users, wearable augmented reality, scalability} 
\end{abstract}

\section{Introduction}
The ability of autonomous vehicles (AVs) to effectively interact with vulnerable road users (VRUs), such as pedestrians, is crucial to ensuring safe operations and public confidence. While pedestrians mainly rely on implicit cues (e.g., motion and motor sounds) from a vehicle to interpret its intention \citep{risto2017human, moore2019case}, explicit signals from a driver, including verbal exchanges, eye contact, and hand gestures, help resolve impasses and instil trust in interactions \citep{rasouli2019autonomous}. Once humans relinquish control to an AV, these informal signals may become less prevalent or possibly disappear altogether. External human–machine interfaces (eHMIs) \citep{dey2020taming} are currently being investigated as a possible way to compensate for the lack of driver cues, allowing for intention transparency, which is a desirable quality in almost every intelligent system \citep{zileli2019towards}.

In order to understand key factors influencing pedestrian behaviour and experiences, most external communication research has evaluated eHMIs in the fundamental traffic setting involving one pedestrian and one vehicle \citep{colley2020unveiling}. However, for eHMIs to become an effective mediator in real-world traffic situations, it is critical for external communication research to take into account scalability factors (i.e., vehicular and pedestrian traffic volumes) and their associated challenges. For example, pedestrians may experience an increased cognitive load when interpreting signals from multiple AVs \citep{mahadevan2018communicating, dey2020taming} or mistakenly believe a message intended for another is directed to them \citep{dey2021towards}.

One promising solution to the scalability issues is incorporating augmented reality (AR). This technology has been explored in the automobile industry to improve driving safety and comfort \citep{riegler2021augmented}. In-car AR, such as heads-up and windshield displays, offer diverse opportunities to aid navigation, highlight potential hazards, and allow for a shared perception between a driver and an automated driving system \citep{wiegand2019incarar}. The application of AR outside of vehicles to assist AV–pedestrian interaction is also of increasing interest in academia \citep{ tabone2021vulnerable}. As with smartphones, the personal nature of wearable AR \footnote{In this paper, we use the term \textit{wearable AR} to refer to all types of near-eye displays regardless of their form factor. These displays include head-mounted AR devices (e.g., Microsoft HoloLens), monocular and binocular AR glasses (e.g., Google Glass), and contact lenses (e.g., Mojo Lens).} allows their connected eHMI concepts to address an unlimited number of road users simultaneously with notable precision and resolution \citep{dey2020taming}. In addition, tailored communication based on user preferences and characteristics may contribute to eHMIs becoming more inclusive. Notably, AR has been investigated as an accessibility tool for visually impaired people \citep{coughlan2017ar4vi}. Most significantly, wearable AR enables digital content to be displayed within the physical environment, allowing users to retain situational awareness and react rapidly to safety alerts \citep{tong2021augmented}. 


Various AR concepts have been designed to convey road-crossing information \citep{hesenius2018don, prattico2021comparing, tabone2021towards} and provide collision warnings to pedestrians \citep{tong2021augmented}. However, to our knowledge, no studies have been undertaken to date to evaluate AR concepts in a complex traffic setting where pedestrians must consider the intentions of several AVs in making crossing decisions. Our driving assumption is that a multi-vehicle situation necessitates the understanding of an appropriate communication approach to provide pedestrians with pertinent cues without overwhelming them. Furthermore, the literature has focused on determining the efficacy of various AR concepts in conveying AV intent rather than pedestrians’ preferences for using wearable AR in daily interactions with AVs. Given the novel AR experiences and the shift away from using public crossing facilities and toward using personal devices, it is important to gauge pedestrians’ acceptance of wearable AR solutions. 


To address these research gaps, we designed three AR eHMI concepts with different ways to signal responses from multiple AVs: a visual cue on each vehicle, a visual cue that represents all vehicles, and both the aforementioned types of visual cues. We used virtual reality (VR) to simulate and test wearable AR prototypes against a pedestrian push button baseline. Our overall research goal was to answer the following research questions: \textbf{(RQ1)} \textit{To what extent, if any, do pedestrians prefer using wearable AR to interact with AVs?} \textbf{(RQ2)} \textit{How do different communication approaches influence pedestrians' perceived cognitive load and trust?}

Our study makes the following contributions: we (1) present novel AR eHMI concepts that assist the crossing of pedestrians in heavy traffic scenarios, (2) identify factors influencing pedestrians’ preferences for wearable AR solutions, and (3) determine the effect of three distinct communication approaches on pedestrians’ crossing experiences.

\section{Related Work}

\subsection{External Communication of Autonomous Vehicles}

The vast majority of car crashes are caused by human error \citep{treat1979tri, hendricks2001relative}; therefore, advanced driver assistance systems have been developed to assist drivers in a variety of driving tasks (e.g., active cruise control, collision warnings) or relieve them fully from driving. Without active drivers, future vehicles may be outfitted with additional interfaces that communicate clearly with pedestrians and other VRUs regarding their intentions and operating states. For instance, Waymo has submitted a patent stating that cars may inform pedestrians using ``a physical signalling device, an electronic sign or lights, [or] a speaker for providing audible notifications'' \citep{urmson2015pedestrian}. Meanwhile, Uber has further proposed using a virtual driver and on-road projections \citep{sweeney2018light}. Potential implementations of eHMI also include approaches in which communication messages are detached from the vehicles. The urban technology firm Umbrellium has prototyped an LED-based road surface capable of dynamically adapting its road markings to different traffic conditions to prioritize pedestrians' safety \citep{umbrellium2017}. In addition, Telstra has trialed a technology enabling vehicles to deliver early-warning collision alerts to pedestrians via a smartphone \citep{cohda2017}. 

The locus of communication – \textit{Vehicle}, \textit{Infrastructure}, and \textit{Personal Device} – is one of the key dimensions in the eHMI design space \citep{colley2020design}. According to a review of 70 different design concepts from industry and academia, vehicle-mounted devices have accounted for the majority of research on the external communication of AVs thus far \citep{dey2020taming}. However, urban infrastructure and personal devices are promising alternatives for facilitating complex interactions involving multiple road users and vehicles due to their high scalability and communication resolution \citep{dey2020taming}. 

\subsection{Scalability}

In the context of AV–pedestrian interaction, scalability refers to the ability of eHMIs to be employed in situations with a large number of vehicles and pedestrians without compromising on efficacy \citep{colley2020unveiling}. In this case, the communication relationship goes beyond the simple one-to-one encounters and includes one-to-many, many-to-one and many-to-many interactions \citep{colley2020towards}. Although scalability research is still in its early stages \citep{colley2020unveiling}, potential scaling limitations of eHMIs, including low communication resolution and information overload, have been noted in several research articles \citep{dey2020taming, dey2021towards, robert2019future}. 

In terms of communication resolution, that is, ``the clarity of whom the message of an eHMI is intended" \citep{dey2020taming}, a message broadcasted to all road users in a vehicle's vicinity, for example, from an on-vehicle LED display, might result in misinterpretation. This issue is particularly apparent when co-located road users have conflicting rights of way \citep{dey2020taming}, which may lead to confusion or even unfortunate outcomes in real-world traffic situations. \citet{dey2021towards} tested four eHMI designs with two pedestrians and observed that non-specific yielding messages increased the participants' willingness to cross even when the vehicle was stopping for another person. To address this possible communication failure, \citet{verstegen2021commdisk} prototyped a 360-degree disk-shaped eHMI featuring eyes and dots that acknowledge the presence of multiple (groups of) pedestrians. Other proposed alternatives include nomadic devices, the personal nature of which inherently enables targeted communication, and smart infrastructures (e.g., responsive road surfaces) \citep{dey2020taming}. However, more research is required to determine user acceptance as well as the (cost-) effectiveness of such solutions. 

Information overload may occur when pedestrians are presented with an excessive number of cues. In the study by \citet{mahadevan2018communicating}, a mixed interface of three explicit cues situated on the automobile, street infrastructure, and a pedestrian's smartphone was viewed as time-consuming and perplexing by many participants. \citet{hesenius2018don} reported a similar finding, where participants disliked the prototype that visualizes safe zones, navigation paths, and vehicle intents simultaneously. In the case of multiple AVs, an increase in the number of external displays was expected to impose a high cognitive load onto pedestrians \citep{robert2019future} and turn street crossing into ``an analytical process" \citep{moore2019case}. According to \citet{colley2020inclusive}, when multiple AVs communicate using auditory messages, pedestrians’ perceived safety and cognitive load improve; however, it is uncertain whether the same observation can be made with visual messages. Our study aims to close this knowledge gap by examining three different approaches to displaying visual responses from multiple AVs. 

\subsection{Wearable AR Concepts}

Globally, smartphone uptake has increased at a very swift pace. Together with advances in short-range communication technologies, the devices have been investigated for their potential to improve pedestrian safety, such as aiding individuals in crossing streets \citep{malik2021determining, hollander2020save} and providing collision alerts \citep{wu2014cars, hussein2016p2v}. Smartphones’ close proximity to users allows them to access reliable positioning data for collision estimations and deliver adaptive communication messages based on users’ current phone activity (e.g., listening to music) \citep{liu2015pofs}.

Wearable AR, as the next wave of computing innovation, has similar advantages to smartphones. However, its ability to combine the virtual and real worlds enables a more compelling and natural display of information and improved retention of situational awareness \citep{azuma2019road}. Context-aware and pervasive AR applications \citep{grubert2016towards} also present an opportunity to aid users in more diverse ways. They are envisioned to become smart assistants that can semantically understand the surrounding environment, monitor the user's current states (e.g., gaze, visual attention), and adjust to their situational needs \citep{starner1997augmented, azuma2019road}. This has led to a growing discussion on the application of wearable AR for AV–pedestrian communication. In a position paper where 16 scientific experts were interviewed, it was partially agreed that wearable AR might resolve scalability issues of AV–VRU interaction \citep{tabone2021vulnerable}. Recent work has explored several different AR eHMI concepts but has yet to examine the scalability aspect. \citet{tong2019augmented} designed an AR interface to warn pedestrians of oncoming vehicles while other studies have presented navigational concepts \citep{hesenius2018don, prattico2021comparing} and theoretically-supported prototypes \citep{tabone2021towards} offering crossing advice. Our study attempts to extend this body of work through an empirically based investigation of wearable AR design concepts in a multi-vehicle situation. 

Currently, various technical issues exist that make it challenging to prototype and evaluate wearable AR interfaces outdoors \citep{billinghurst2021grand}: (1) a narrow field of view (FOV) that covers only a portion of the human field of vision, limiting what users can see to a small window; (2) an unstable tracking system that is affected by a wide range of environmental factors (e.g., lighting, temperature, and movement in space); and (3) low visibility of the holograms in direct sunlight. For these reasons, we utilized VR simulations to overcome the shortcomings of wearable AR and the limitations of AV testing in the real world, following a similar approach to~\citet{prattico2021comparing}.

\section{Design Process}
\subsection{Crossing Scenario}
Similar to most studies on AV–pedestrian interaction, we selected an ambiguous traffic situation, that is, a midblock location without marked crosswalks or traffic signals, requiring pedestrians to cross with caution and be vigilant of oncoming vehicles. To assess the design concept's scalability, the crossing scenario featured many vehicles driving in both directions on a two-way street. This situation is prevalent in urban traffic, typically requiring pedestrians to estimate the time-to-arrival of vehicles and select a safe gap to cross. However, the ability to correctly assess the speed and distance of approaching cars varies with different environmental conditions and across demographic groups \citep{rasouli2019autonomous}. Inaccurate judgments may lead to unsafe crossing decisions, causing pedestrian conflicts with vehicular traffic. On the premise that not all road users can chart their best course of action, we sought to create a design concept to aid their crossing decisions.

\subsection{Design Concepts}
\begin{figure}[t]
\begin{center}
\includegraphics[width=\textwidth]{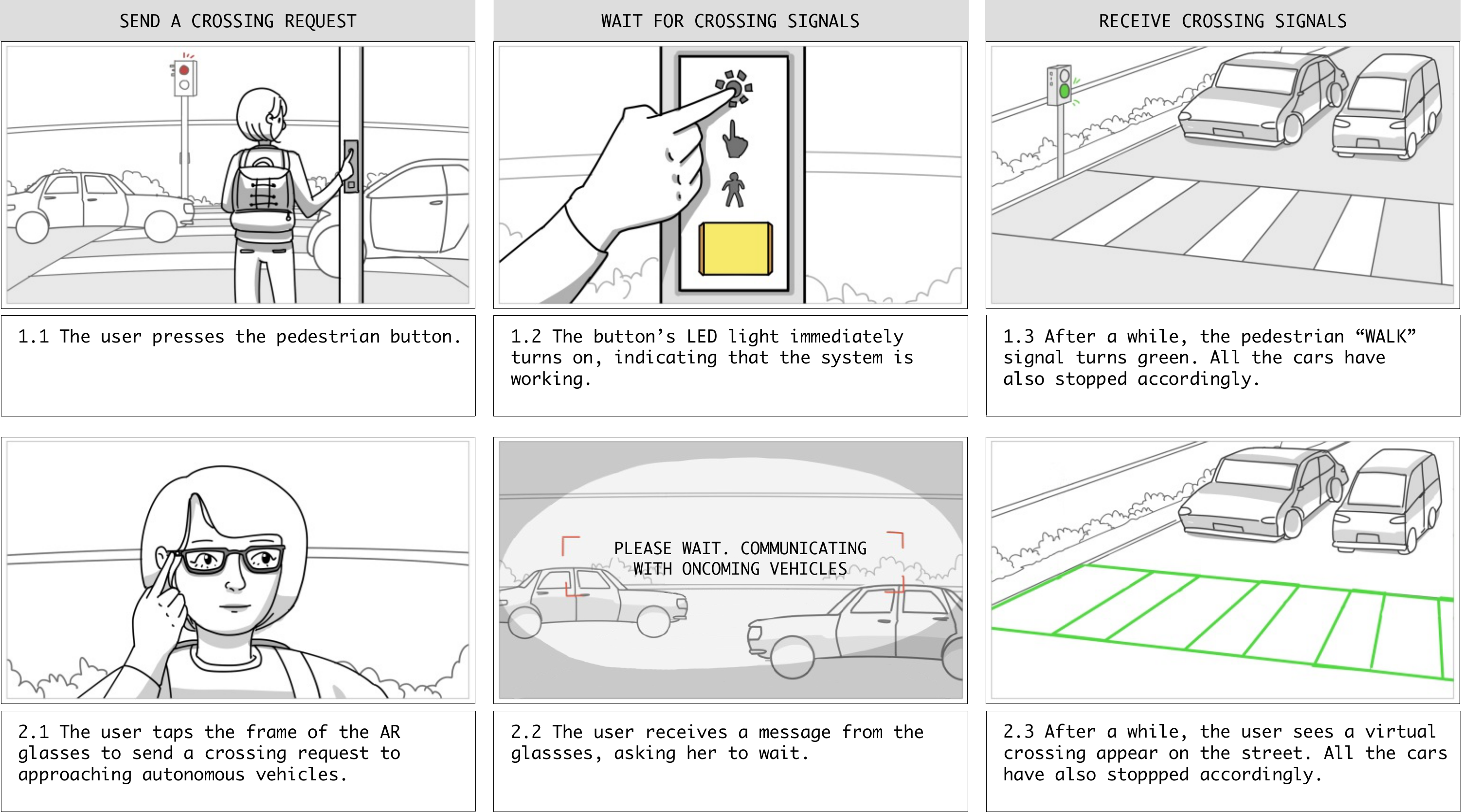}
\end{center}
\caption{Storyboard illustrating three-stage user interactions of the pedestrian button (top) and AR glasses (bottom). Only visual signals were depicted to keep the storyboard simple.}\label{fig:storyboard}
\end{figure}

Our wearable AR concepts were inspired by the widely used pedestrian push button, which enables pedestrians to request a crossing phase. The buttons are typically installed at locations with intermittent pedestrian volumes, where an automatic pedestrian walk phase has not been implemented. The installment is intended to improve vehicle mobility by reducing unnecessary waiting times \citep{lee2013evaluation} and promote pedestrian compliance at traffic signals \citep{van2006pedestrian}. Moreover, accessible push buttons that incorporate audio-tactile signals may be especially beneficial to blind and vision-impaired pedestrians \citep{barlow2005crossroads}.
In the advent of autonomous driving, the pedestrian push button remains an effective solution to mediate conflicts and improve pedestrian safety; however, the system may not be available at every intersection and midblock location. Furthermore, pedestrians tend to cross at convenient locations that present shorter delays \citep{ravishankar2018pedestrian}. Therefore, we followed an iterative design process to devise a concept where pedestrians can utilize the AR glasses to negotiate a crossing opportunity with approaching AVs. Prototypes of varying fidelities were created and improved through internal discussions among the authors. Additionally, two pilot studies (with a total of four participants) were conducted prior to the main investigation. User interactions were modeled after those used with the pedestrian button, comprising three stages, as illustrated in Figure \ref{fig:storyboard} and described in greater detail as follows.

\textit{Sending a crossing request:} As a safety prerequisite, predicting pedestrian crossing intentions based on parameters such as the pedestrian dynamics, physical surroundings, and contextual scene information is one of the most critical tasks of AVs \citep{ridel2018literature}. However, many challenges remain to be overcome in achieving a reliable and robust solution. For this reason, we implemented a user-initiated communication approach with pedestrians explicitly indicating their crossing intents for a greater sense of control. Users can send a crossing request to all nearby AVs by quickly tapping a touch surface on the temples of AR glasses. While various methods for controlling the AR glasses exist, the tapping gesture was selected for its simplicity and ease of prototyping. Additionally, it is widely employed in wireless earphones, smart glasses and smart eyewear (e.g., Ray-Ban Stories).

\textit{Waiting for crossing signals:} Analogous to how the pedestrian push buttons offering visual and audible feedback when pressed, the AR glasses displayed a text prompt acknowledging the crossing request. According to media reports on push-button usage in the United States, many people are unsure if the system is of value and even regard them as placebo buttons, with their presence only offering an ``illusion of control" \citep{prisco2018}. The confusion has arisen mainly because the push buttons are inoperative during off-peak hours or have been supplanted by more advanced systems (e.g., traffic sensors) and kept only for accessible features \citep{prisco2018}. Considering these user frustrations stemming from a lack of understanding regarding how a system works, we ensured that the text prompt briefly explains the workings of the AR glasses.

\textit{Receiving crossing signals:} We developed three communication approaches to visually convey AVs' responses to the crossing request. The first approach involves placing a visual cue on each vehicle (``distributed response"); specifically, the AR glasses render a green overlay that covers a vehicle’s surface to indicate a yielding intent. The idea of an overlay is based on the futuristic digital paints that may be incorporated in automobiles by 2050 \citep{autotrader2020}. Given the lack of consensus regarding the optimal placement of visual cues on a vehicle's body, an overlay offers the advantage of being noticeable and visible from various angles. Green was chosen as the color to indicate ``go" because of its intuitiveness \citep{dey2020color}; we also assumed that possible confusion in perspectives \citep{bazilinskyy2019survey} is less likely to occur when the user initiates the communication. The second approach entails the use of a single visual cue, in this case, an animated forward-moving pedestrian crossing, to convey the intentions of all cars (``aggregated response"). The zebra crossing is a widely recognized traffic symbol that numerous eHMI studies have investigated \citep{nguyen2019designing, locken2019should, prattico2021comparing, tabone2021towards, dey2021towards}; its forward movement indicates the crossing direction \citep{nguyen2019designing}, and the markings have high visibility \citep{locken2019should}. The third approach combines both types of visual cues by displaying car overlays and an animated zebra crossing simultaneously. This approach was implemented based on study findings from~\citet{hesenius2018don}, taking into account the possibility of participants having different preferences regarding different combinations of cues.

\section{Evaluation Study}
\subsection{Study Design}
Given that the AR eHMI concepts were designed for a multi-vehicle traffic situation, a comparison to a currently implemented system would yield relevant insights into pedestrian preferences and crossing experiences. Therefore, we decided on a within-subjects study design with four experimental conditions: a baseline pedestrian push button and three wearable AR concepts with different communication approaches - aggregated response (AR crosswalk), distributed response (AR overlay), and both the aforementioned types (see Figure \ref{fig:design-concepts}). To minimize carryover effects, we changed the order of presenting the concepts from one participant to another using a balanced Latin Square. We kept factors that might influence pedestrian behavior, such as vehicle speed, deceleration rates, and gaps between vehicles the same across all conditions. The participants' experimental task was to stand on the sidewalk, several steps away from traffic, and cross the street with the assistance of a given design concept.

\subsection{Participants}
To determine the required sample size, an a priori power analysis was performed using G*Power \citep{faul2009statistical}. With an alpha level of .05, a sample of 24 participants was adequate to detect medium effect sizes (Pearson's r = .25) with a power of .81 \citep{cohen2013statistical} for our measures. 

We recruited 24 participants (62.5\% female; 18-34 age range) through social media networks and word of mouth. The participants included working professionals and university students who had been living in the current city for at least 1 year and who could speak English fluently. All participants were required to have normal or corrected-to-normal eyesight, as well as no mobility impairment. Of our participants, 13 had tried VR a few times, and three had extensive experience with it. Meanwhile, only two participants reported having experienced AR once. Thirteen participants required prescription \hl{g}lasses; the remaining 11 had normal visual acuity, three of which had undergone laser eye surgery, and one was using orthokeratology (i.e., corneal reshaping therapy) to correct their vision. The study was conducted at a shared workspace in Ho Chi Minh City (Vietnam), following the ethical approval granted by the University of Sydney (ID 2020/779). Participants in this study did not receive any compensation.

\begin{figure}[t]
\begin{center}
\includegraphics[width=\textwidth]{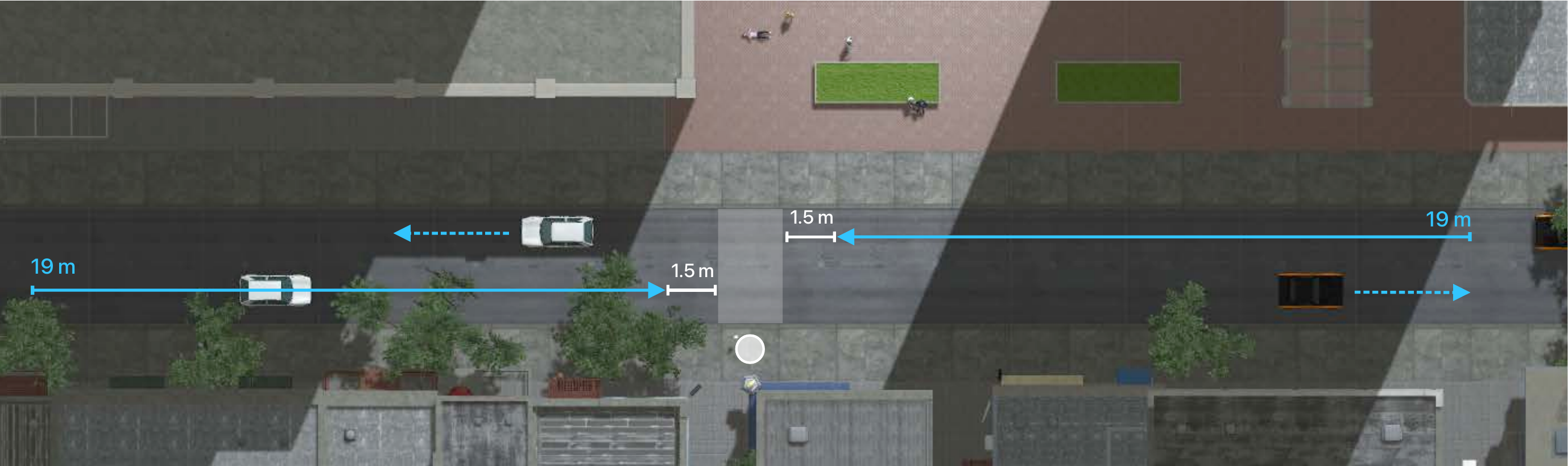}
\end{center}
\caption{A top-down view of the virtual environment zooms in on the midblock location where participants made the crossing. Dotted blue arrows indicate the travel direction of AVs. Solid blue arrows indicate where the AVs (on each lane) begin to decelerate and where they come to a complete halt. White circle indicates pedestrian position at the start of each experimental condition.}\label{fig:topview}
\end{figure}

\subsection{Virtual Reality Prototype}
\textit{Apparatus}. The VR prototype was developed using the Unity \footnote{https://unity.com/} game engine and experienced with the Oculus Quest 2 VR system \footnote{https://www.oculus.com/quest-2/}. The head-mounted display (HMD) provides a fully untethered 6DOF experience and hand tracking feature, allowing users to walk around freely and engage in VR naturally with their hands. The experiment was conducted in an 8x3-meter open floor space, where participants were able to physically walk two-thirds of the street before being teleported to the other side. The (auto) teleportation was used to overcome HMD tracking space limits and to ensure that participants could observe how the visual cues disappeared and the AVs resumed driving after their crossing.

\textit{Virtual environment}. The virtual environment was modeled using commercially available off-the-shelf assets. The scene featured an unmarked midblock location on a two-way urban street. Pedestrian crossing facilities, including traffic lights and zebra crossings, were only available under the experimental condition where pedestrians crossed the street using the pedestrian push button. To create a more realistic social atmosphere, we used Mixamo 3D characters \footnote{https://www.mixamo.com/} to replicate human activities on the sidewalk: some individuals were exercising while others were speaking with one another. Additionally, an urban soundscape with bird chirping sounds and traffic noise was included.

The vehicles used in this experiment were obtained from the Unity Asset Store and comprised a black/orange sports car, a silver sedan, and a white hatchback to create a more natural perception of traffic. Despite their model differences, these vehicles had similar sizes and kinematic characteristics, both of which were found to influence pedestrian experience and behavior~\cite{dey2017impact}. Vehicular traffic was composed of fully automated cars (Level 5) \citep{sae2021taxonomy} traveling in both lanes. To create the perception of autonomous driving, we did not model people inside and implemented a futuristic Audi e-tron sound \footnote{https://www.e-tron-gt.audi/en/e-sound-13626} for each vehicle. The number of vehicles in each lane varied, but they consistently travelled with impassable gaps to ensure that participants could not cross without the AVs yielding. In the simulation, the vehicles were spawned at a location hidden from the participants' view; they started accelerating and driving at approximately 30 km/h before making a right turn. When responding to pedestrians' crossing requests, the vehicles slowed down at a distance of 19 m, following the safe stopping distance recommended in urban zones \footnote{https://roadsafety.transport.nsw.gov.au/speeding/index.html}). They came to a complete stop at 1.5 m from the designated crossing area and only resumed driving once the participants had reached the other side of the road (see Figure \ref{fig:topview}).

\begin{figure}[t]
\begin{center}
\includegraphics[width=\textwidth]{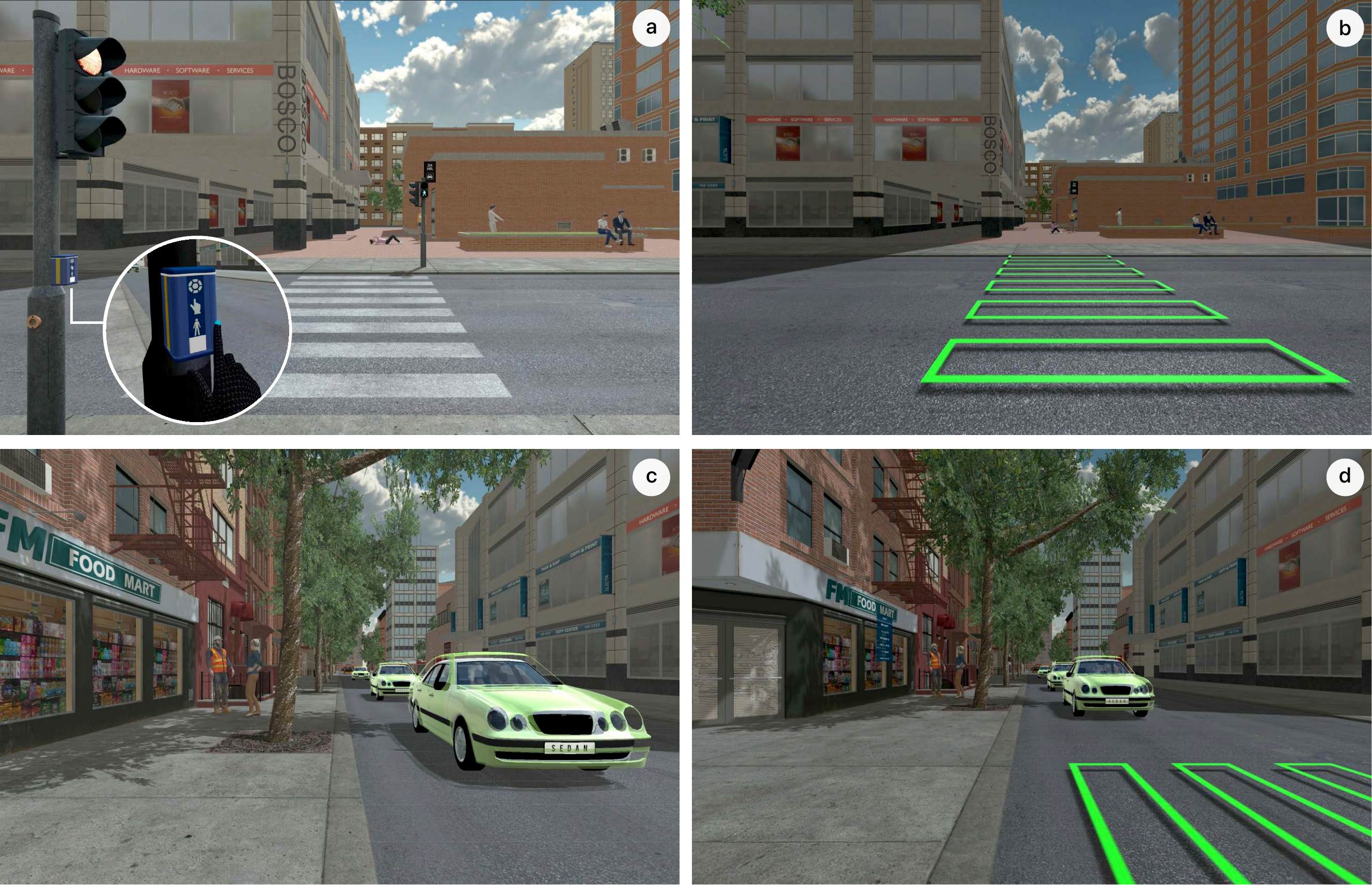}
\end{center}
\caption{Simulation environment and interfaces included in the evaluation: (a) Pedestrian push button; (b) AR crosswalk; (c) AR overlay; and (d) AR-combined.}\label{fig:design-concepts}
\end{figure}

\textit{Evaluated concepts}. We commissioned a game artist to create a 3D model of the Prisma TS-903 button \footnote{https://www.prismatibro.se/en/prisma-ts-903-eng/} used in the city where the study took place. In VR, participants could use their hands to engage with the button in the same way they would in real life (see Figure \ref{fig:design-concepts}a). To experience the wearable AR concepts, participants used the VR headset as if it was a pair of AR glasses and were instructed to tap on its side whenever they planned to cross. On Oculus Quest 2, this type of tap gesture was not available; it was prototyped by creating an invisible collision zone around the HMD that detects any contact with the user's fingertips. The tapping immediately triggers sound feedback and displays a HUD text prompt \textit{``Please wait. Communicating with oncoming vehicles"} in users' primary field of vision. After 9 s, all AVs responded to the pedestrian crossing request by decelerating at a distance of 19 m and displaying the car overlays. However, those with a short stopping distance (already approaching the pedestrians when the request was received) continued to drive past to avoid harsh braking. To account for these cars, a 3-second delay was put in place to make sure that the AR zebra crossing only appeared when the crossing area was safe. The design of the zebra crossing was inspired by the Mercedes-Benz F 015 concept \footnote{https://www.mercedes-benz.com/en/innovation/autonomous/research-vehicle-f-015-luxury-in-motion/}, with bright neon green lines and flowing animation (see Figure \ref{fig:design-concepts}b). The car overlay was made of semi-transparent emissive green texture and appeared to be a separate layer from the car (see Figure \ref{fig:design-concepts}c). Both the zebra crossing and the car overlay are conformal AR graphics situated as parts of the real world. In addition to visual cues, we offered audible signals to indicate wait time (slow chirps) and crossing time (rapid tick-tock-tick-tock). These sounds are part of the Australian PB/5 push button signaling system \footnote{https://www.maas.museum/inside-the-collection/2010/04/16/pedestrian-button-1980s-australian-product-design-pt2/}, and they were implemented across four experimental conditions. 


\subsection{Procedures}

After the participants had signed up for the study, a screening questionnaire was used to obtain their demographic information, including age group, gender, English proficiency level, occupation, nationality, length of stay in the current city, walking issues, and eye conditions. On the day of the study, we welcomed the participants and gave them a brief overview of the study and the related tasks. The participants were then asked to read and sign a consent form. Following a quick introduction to the VR system, we asked the participants to put on the HMD and adjust it until they felt comfortable and could see the virtual environment clearly. A glasses spacer was inserted in the HMD such that the participants could wear the headset with their glasses on. 

Before beginning the experiment, the participants took part in a familiarization session in which they practiced crossing the street and interacting with virtual objects with their hands. Prior to each experimental condition, we presented the participants with an image of the pedestrian push button or the AR glasses to gauge their familiarity with the technology and inform them about the system with which they would be engaging. However, they were not made aware of the differences between the wearable AR concepts. After each condition, the participants removed their headsets and completed a series of standardized questionnaires at a nearby table. We also ensured that no participant was experiencing motion sickness and that all could continue with the experiment. After all the conditions had been completed, we conducted a semi-structured interview to gain insights into their experiences. 
\begin{figure}[t]
\begin{center}
\includegraphics[width=\textwidth]{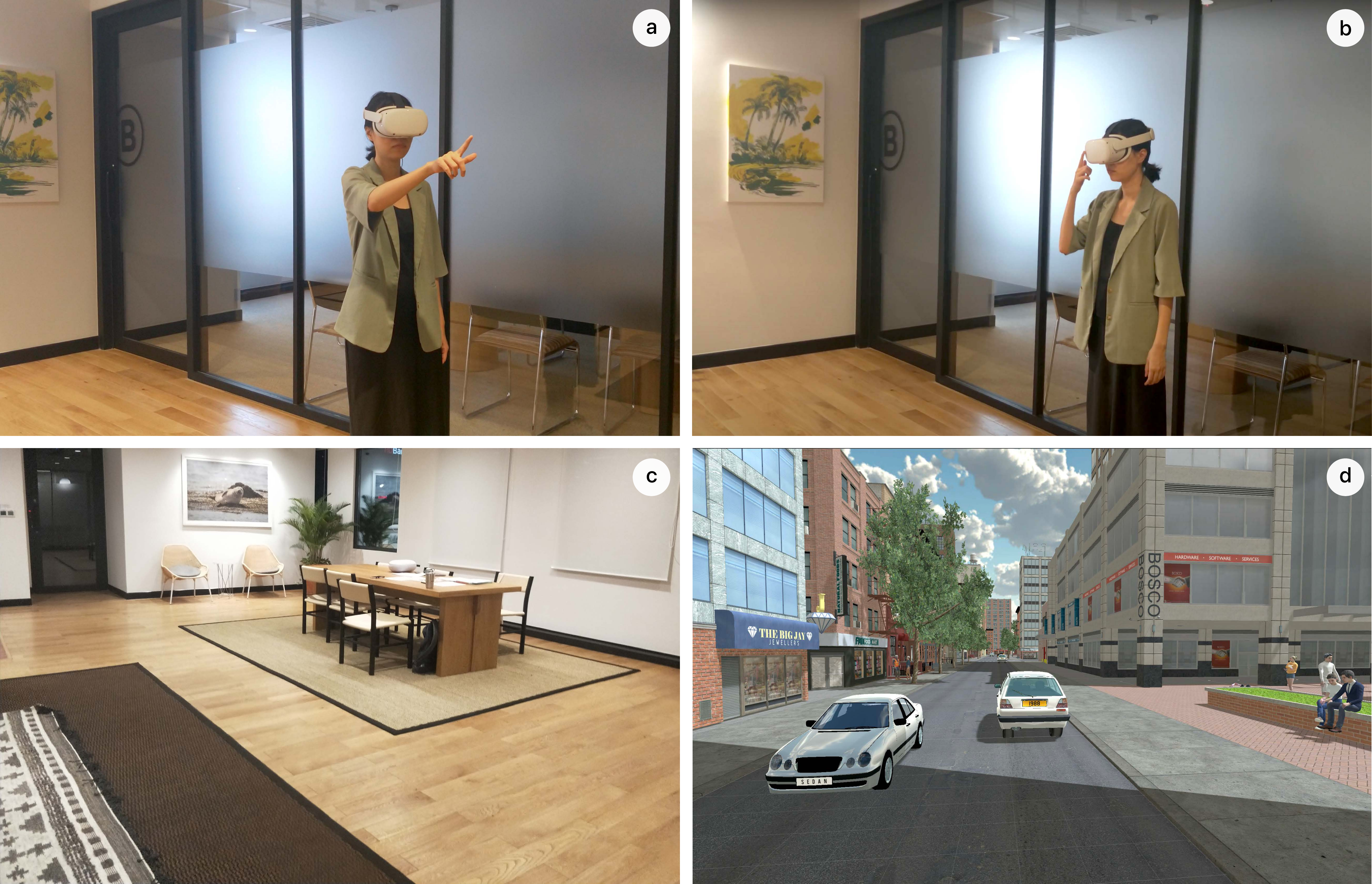}
\end{center}
\caption{Experimental setup: (a) The participant pressing the (virtual) pedestrian button; (b) the participant tapping the side of the HMD; (c) walking space and interview table; and (d) virtual environment}\label{fig:experiment-setup}
\end{figure}

\subsection{Data Collection}


After each experimental condition, we monitored participants' simulator sickness with the single-item Misery Scale \citep{bos2010effect}. If the rating was four or higher, the study would be suspended. We then measured perceived cognitive load with the NASA Task Load Index (NASA-TLX) \citep{hart1988development} on a 20-point scale. The questionnaire has six workload-related dimensions: Mental Demand, Physical Demand, Temporal Demand, Performance, Effort, and Frustration. These dimensions were combined into one general cognitive load scale Cronbach’s $\alpha = .783$). To assess trust in human-machine systems, we used a 12-item trust scale \citep{jian2000foundations}. The first five items provided an overall distrust score (Cronbach’s $\alpha = .896$); the next seven items provided an overall trust score (Cronbach’s $\alpha = .941$). Lastly, the 10-item System Usability Scale \citep{brooke1996sus} was used to measure usability (Cronbach’s $\alpha = .909$). All the questionnaires were explained to the participants and administered under supervision. We also instructed participants to assess the prototyped systems instead of the VR representation.

After the completion of all experimental conditions, the participants were asked to rank the systems from 1 (most preferred) to 4 (least preferred). Additionally, a semi-structured interview was conducted to gain a better understanding of their overall experience, the reasoning behind their preferences, and their perspectives on various system aspects and the VR simulation. 

\subsection{Data Analysis}
\textit{Questionnaires}: We first calculated summary statistics and created data plots to investigate the data sets. We assessed the normality of data using Shapiro-Wilk tests and a visual inspection of their Q-Q plots. Because most data have non-normal distribution, we used the non-parametric Friedman test to determine any statistically significant differences in questionnaire outcomes. In case of significant differences, we performed Dunn–Bonferroni procedure for multiple pairwise comparisons as post-hoc tests. We considered an effect to be significant if $p < .05$. IPM SPSS version 28 was used for all statistical analyses. 

\textit{Interviews}: Post-study interviews were transcribed by the interviewer with the assistance of an AI-based transcription tool. Two coders performed an inductive thematic analysis \citep{braun2006using} to identify and interpret patterns (themes) within the data. The first coder (TT) had extensive knowledge of the study, while the second coder (YW) was not involved in its conception and implementation. This approach enabled us to have a more complete and unbiased look at the data. 

The analysis began with the first coder selecting a subset of six interviews (25\% data units) with good representativeness. The first round of coding was performed independently by both coders, followed by a discussion to agree on the coding frame. In the second round, the first coder applied the coding frame to all interviews. Finally, we examined the themes and patterns that emerged, which composed part of the Results section. 

\section{Results}

\subsection{Concept Ranking}

Regarding the top preference (first ranking), approximately half of the participants preferred the AR concept incorporating both the animated crosswalk and car overlays, while one-third favored the pedestrian button. The \textit{AR overlay}, followed by the \textit{AR crosswalk}, were the least preferred (see Table \ref{tab:ranking}).

A Friedman test showed a significant difference in the mean rankings among concepts ($\chi^2(3) = 29.850, p < .001$). Post-hoc tests revealed that the \textit{AR-combined} ($mdn=1.0$) was rated significantly higher than the \textit{AR crosswalk} ($mdn=3.0$) ($z = 1.375, p_{corrected} = .001$) and \textit{AR overlay} ($mdn=4.0$) ($z = 1.875, p_{corrected} = .000$), but not the pedestrian button ($p_{corrected} = .705$). The pedestrian button ($mdn=2.0$) was rated significantly higher than the \textit{AR overlay} ($mdn=4.0$) ($z = -1.292, p_{corrected} = .003$).


\begin{table}[h!]
  \caption{Ranking results by frequency of nomination.}~\label{tab:ranking}
  \centering
  \begin{tabular}{l l l l l}
  \toprule
        & {\small{Button}}
        & {\small {AR crosswalk}}
        & {\small {AR overlay}}
        & {\small {AR-combined}} \\
    \midrule
        \vspace{0.2cm}
        1st rank & 8 & 2 & 1 & 13  \\
        \vspace{0.2cm}
        2nd rank & 8 & 6 & 1 & 9  \\
        \vspace{0.2cm}
        3rd rank & 5 & 8 & 9 & 2  \\
        4th rank & 3 & 8 & 13 & 0 \\
    \bottomrule
  \end{tabular}
\end{table}

\begin{table}[h!]
  \centering
  \caption{Mean and standard deviations for NASA-TLX scores, SUS scores, and Trust Scale ratings}
  \begin{tabular}{l l l l l}
  \toprule
    & {\small{Button}}
    & {\small {AR crosswalk}}
      & {\small {AR overlay}}
      & {\small {AR-combined}} \\
     & (M / SD) & (M / SD) & (M / SD) & (M / SD) \\
    \midrule
    \vspace{0.2cm}
    SUS & \textbf{85.31} / 12.30 & \textbf{77.29} / 19.78 & \textbf{74.48} / 20.61 & \textbf{85.10} / 13.58\\
    \vspace{0.2cm}
    NASA-TLX & \textbf{20.94} / 13.87 & \textbf{20.90} / 14.55 & \textbf{21.35} / 14.23 & \textbf{13.99} / 8.53\\
    \vspace{0.2cm}
    Trust (subscale) & \textbf{5.83} / .96 & \textbf{5.02} / 1.39 & \textbf{4.82} / 1.39 & \textbf{5.52} / 1.17\\
    Distrust (subscale) & \textbf{1.85} / 1.10 & \textbf{2.32} / 1.21 & \textbf{2.60} / 1.45 & \textbf{1.99} / .94\\
    \bottomrule
  \end{tabular}
  \label{tab:meansd}
\end{table}

\subsection{SUS}
\begin{figure}[h!]
\begin{center}
\includegraphics[width=8.5cm]{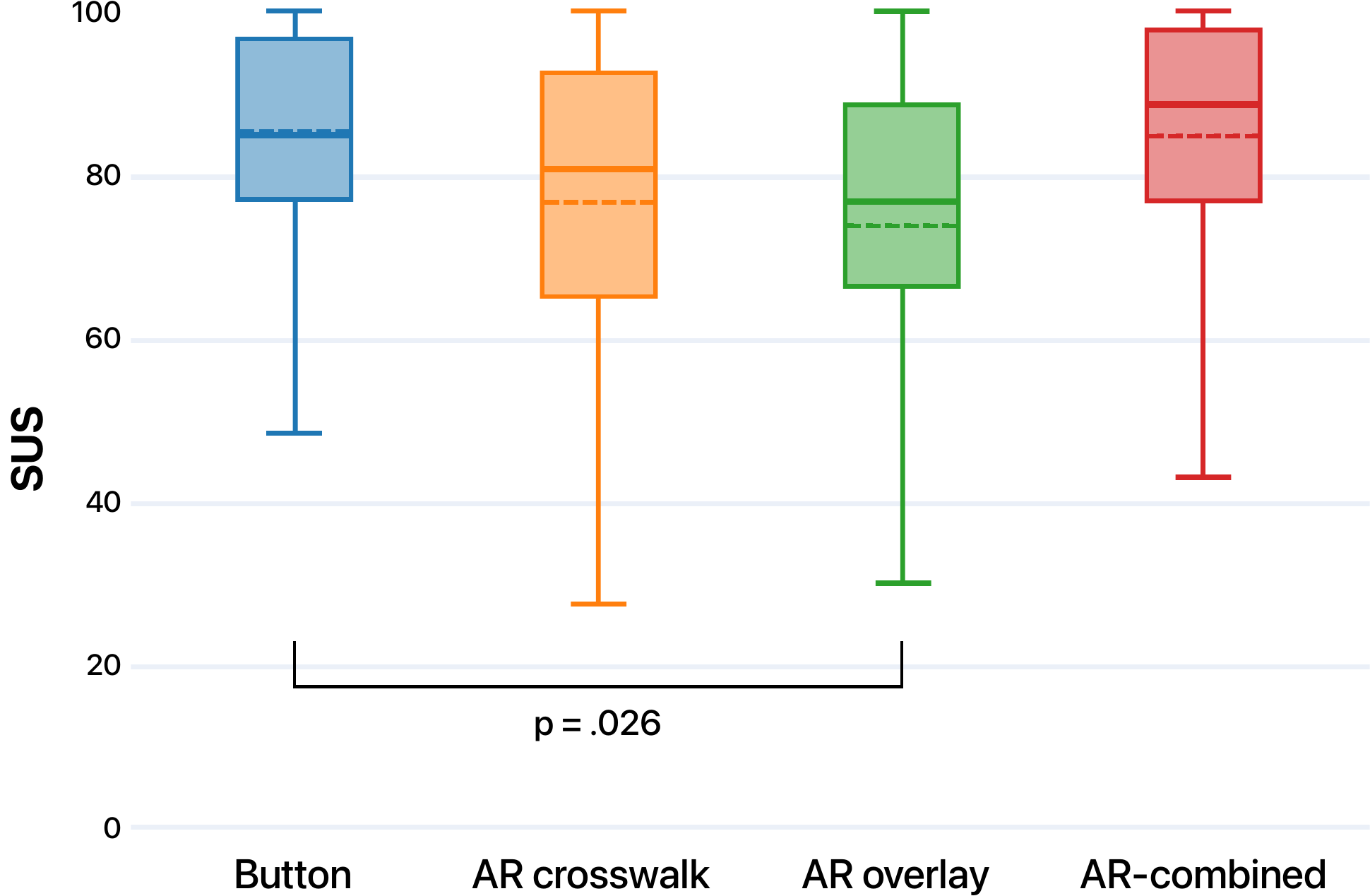}
\end{center}
\caption{Results of the SUS questionnaire. Median = solid line; mean = dotted line, p-values reported for significant pairwise comparisons.}\label{fig:sus}
\end{figure}

Based on the grade rankings created by \citet{bangor2009determining}, the SUS scores of the \textit{Button} and the \textit{AR-combined} were considered as ``excellent". The \textit{AR crosswalk} and the \textit{AR overlay} had lower scores which were in the ``good" range (see Table \ref{tab:meansd}). A Friedman test indicated a significant main effect of the concepts on the usability scores ($\chi^2(3) = 10.808, p = .013$). Post-hoc analysis revealed that the usability scores were statistically significantly different between the \textit{Button} ($mdn=85$) and the \textit{AR overlay} ($mdn=77.50$),  ($z = 1.063, p_{corrected}=.026$), as shown in Figure \ref{fig:sus}.

\subsection{NASA-TLX}
\begin{figure}[h!]
\begin{center}
\includegraphics[width=8.5cm]{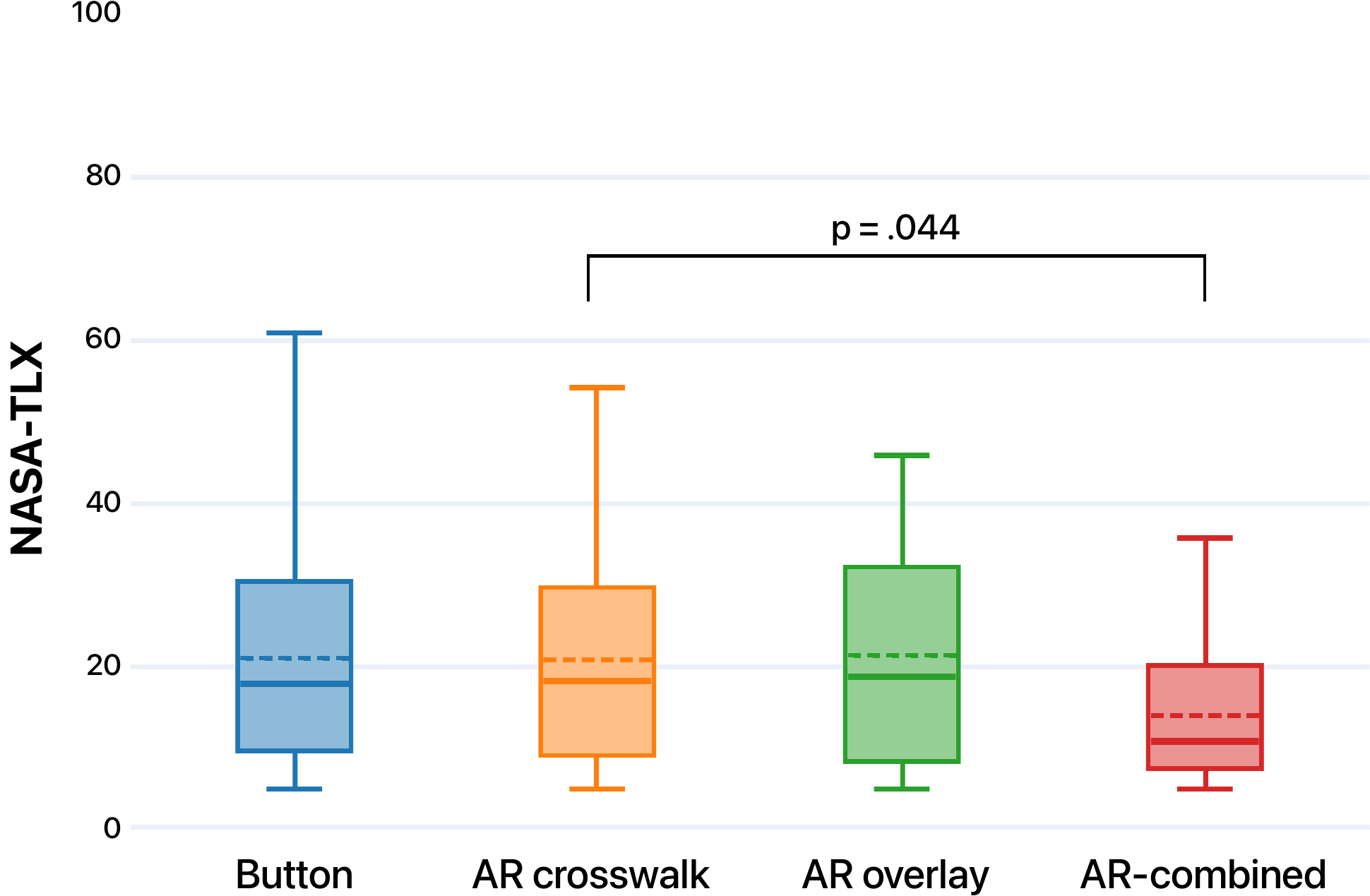}
\end{center}
\caption{Results of the NASA-TLX questionnaire. Median = solid line; mean = dotted line, p-values reported for significant pairwise comparisons.}\label{fig:nasa}
\end{figure}

Descriptive data analysis showed that the overall scores were low for all concepts; however, the \textit{AR-combined} elicited the least cognitive load (see Table \ref{tab:meansd}). A Friedman test showed a significant difference in the overall mean scores ($\chi^2(3) = 11.535, p = .009$). Post-hoc tests revealed that the \textit{AR crosswalk} received significantly higher cognitive load scores ($mdn=18.33$) compared to the \textit{AR-combined} ($mdn=10.84$) ($z = 1.000, p_{corrected}=.044$), as shown in Figure \ref{fig:nasa}.

Regarding subscales, the Friedman test found a statistically significant effect of the concepts on \textit{temporal demand} ($\chi^2(3) = 12.426, p = .006$) and \textit{frustration} ($\chi^2(3) = 8.392, p = .039$). The post-hoc tests showed no significant differences ($p_{corrected} > .05$). However, the uncorrected p-values indicated that the \textit{AR-combined} received significantly lower scores in \textit{temporal demand} compared to all other concepts. 


\subsection{Trust Scale}

\begin{figure}[h!]
\begin{center}
\includegraphics[width=\textwidth]{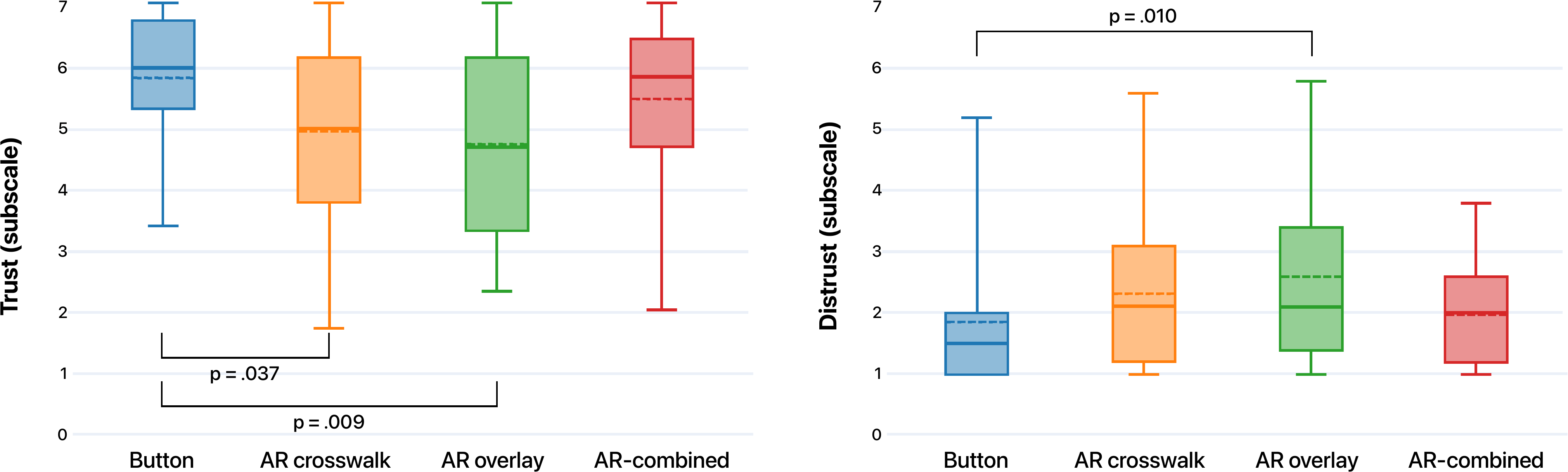}
\end{center}
\caption{Results of the Trust Scale questionnaire: trust subscale (left) and distrust subscale (right). Median = solid line; mean = dotted line, p-values reported for significant pairwise comparisons.}\label{fig:trust-scale}
\end{figure}

According to descriptive data analysis (see Table \ref{tab:meansd}), the participants’ trust in the three AR concepts was lower than in the \textit{Button}, with the lowest trust in the \textit{AR overlay}. Results from a Friedman test found a significant difference in the mean scores of trust ratings ($\chi^2(3) = 14.724, p = .002$). Post-hoc tests revealed that the \textit{Button} ($mdn=6.00$) received significantly higher trust ratings compared to the \textit{AR crosswalk} ($mdn=5.07$) ($z = 1.021, p_{corrected}=.037$) and the \textit{AR overlay} ($mdn=4.79$) ($z = 1.188, p_{corrected}=.009$), as shown in Figure \ref{fig:trust-scale} on the left.

Participants' distrust in the three AR concepts, conversely, was higher than that in the \textit{Button}, with the strongest level of distrust being shown in the \textit{AR overlay} (see Table \ref{tab:meansd}). A Friedman’s test showed a significant difference in the mean ratings ($\chi^2(3) = 15.556, p = .001$). Post-hoc tests revealed that the \textit{AR overlay} ($mdn=2.10$) received significantly higher distrust ratings compared to the the \textit{Button} ($mdn=1.50$) ($z = -1.167, p_{corrected}=.010$), as shown in Figure \ref{fig:trust-scale} on the right.

\subsection{Qualitative Feedback}

This section presents the primary themes that emerged from our qualitative data analysis, providing insight into the participants' perceptions of wearable AR concepts and the design features that influenced their experiences.

\textbf{Wearable AR concepts were unfamiliar yet exciting}. The post-study interviews showed that the participants' familiarity with the design solutions appeared to influence their trust in them. The pedestrian push button was perceived as highly familiar by a noticeable ratio of the participants (n=10). This sense of familiarity was often linked to past experiences (n=8) and had frequently resulted in feelings of confidence while crossing (n=7). For example, P23 stated, \textit{`I feel safer because it's something that I'm used to. I have the feeling that it's guaranteed'}. Wearable AR applications, on the other hand, were regarded as novel and less familiar than the traditional infrastructure (n=6), which might hinder their uptake, especially in the older generation (P4 and P20). As a result, several participants recommended that providing onboarding tutorials (P7) or a user manual (P2) might benefit their adoption. Furthermore, a number of participants stated that additional exposure to wearable AR applications is necessary to establish their dependability (n=7) – \textit{`I have only experienced it once. Maybe I need to interact and use it a few times. I need to try it more to know if it's reliable'} (P21). P7 added that knowledge of relevant statistics, such as the number of users of the AR system, could also contribute to an increase in trust.


Despite the unfamiliarity, wearable AR solutions were frequently described as exciting and cool (n=5). As commented by P2, \textit{`It's like I have mind control and being able to stop all the cars'}. In contrast, the pedestrian push button was deemed to be a conventional system to support pedestrian crossing (n=4), referred to as \textit{`very old school'} (P16) and \textit{`less technologically advanced'} (P13). 


\textbf{Wearable AR offered both advantages and disadvantages as a personal device}. The pedestrian push button baseline enabled a direct comparison of a solution based on personal devices with an infrastructure-based solution, producing a variety of insightful perspectives from the participants on the personal nature of AR eHMIs. The analysis showed that one of the most commonly noted advantages of wearable AR concepts is the increased flexibility of crossing locations (n=4), as opposed to the fixed installation of the pedestrian push buttons. P7 found it particularly useful when \textit{`[she] wants to cross the street in a hurry'} (P7). Furthermore, P5 noted that the precision of requests sent to the vehicles could result in higher efficiency – \textit{`normal vehicles usually focus on the street; maybe they will miss my request to cross the street. If I use the AR glasses, it'd be quicker I think'}. Nonetheless, cost (n=4) and data privacy (n=4) were identified as two of the most significant barriers to personal devices being adopted over public infrastructure. Two participants also raised concerns about circumstances where they might forget the personal device at home (P6 and P9) or do not wish to wear the AR glasses at times (P9). Likewise, personal devices were perceived to be inferior to public infrastructure in terms of liability (n=3) and maintenance (n=2). As commented by P8, \textit{`because the button is of the government, if there's something happened, we can find somebody to blame'}, while P4 stated that \textit{`[she] believe[s] there will be someone taking good care of a public system'}.



The physical form factor of the AR glasses was found to influence their acceptance. The idea of wearing glasses (or even contact lenses) was not appealing to individuals who had undergone eye corrective surgery. The concern was less about the aesthetic qualities of the AR glasses but more about the (re)dependence on eyewear on a daily basis (n=3). Three participants questioned the necessity of using AR glasses to aid in crossing. Further, two people suggested smartphones (P1), smartwatches (P8), and AVs' pedestrian detection feature (P8) as alternative solutions to wearing AR glasses. Nonetheless, three participants identified the potential of using AR glasses for multiple purposes, such as reading the news and watching television (P9), rather than solely assisting in crossing.


\textbf{User-initiated communication provided a sense of control}. When questioned about the preferred mode of interaction, all participants (n=24) responded that they favored sending a crossing request to AVs rather than waiting for the vehicles to detect their intentions. We found that the participants' reasoning regarding this preference revolved around two aspects. First, some participants were skeptical about the ability of AVs to capture intricate human intentions (n=13), stating that pedestrians may cross the street \textit{`spontaneously'} (P12),  \textit{`change their minds'} quickly, or move in ways that suggest something unintentionally (P16). One participant doubted the reliability of algorithms that learn from \textit{`previously fed data’} (P13). Second, some participants preferred to have some control over the interaction (n=9); according to them, proactive communication with AVs was deemed critical for ensuring accuracy and hence safety (n=9). Concerns about the passivity and uncertainty associated with waiting were also mentioned: \textit{`I have no way to know that whether they will stop or not. What if [the cars] just keep moving?'} (P21). 

The qualitative analysis further suggested that the participants preferred a digital approach over bodily gestures (e.g., waving hands), owing to the lack of confidence that AVs would all be able to observe their signals (n=7). For example, P15 stated, \textit{`If I raise my hand, I’m not sure if all cars see it.'} Nonetheless, whereas the integration with traffic lights enables the pedestrian buttons to operate effectively in mixed traffic situations, the practicality of wearable AR to communicate with manual vehicles (n=5) and the extent to which human drivers cooperate (n=6) were questioned. Furthermore, six participants expressed reservations about potential traffic disruptions in the event of many road users using the AR glasses for street crossing. P14 stated, \textit{`what if there were 10, 20 people wearing glasses, but they do not cross the street at the same time?’}.



\textbf{Clear communication mechanism with AVs influenced the perceived safety}. We found that the perceived connection between the system used and AVs influenced the participants' feeling of safety. Regarding the AR glasses, the connection was seen as direct and explicit (n=7). The provision of visual cues assured the participants that the connection was \textit{`established'} and would continue to be maintained during their crossing, as reported by P1: \textit{` I assumed that the vehicles would be waiting for me to finish the crossing. They will allow me as much as possible time to cross […] because they may be connected to my glasses and aware of my presence'}. In the case of the pedestrian button, user feedback revealed divided viewpoints. Eight participants were puzzled as to how the system {``talked"} to the vehicles. P21 thought that \textit{`the digital context [was] missing}, while P17 viewed the two entities as \textit{`disconnected'}. Meanwhile, nine participants contended that the AVs came to a halt due to a changing traffic signal. P18 highlighted that the vehicles \textit{`might have a sensor to read the colour of the traffic light'}. It was this interpretation and confidence in the ability of the traffic lights to regulate traffic that allowed these participants to feel more at ease in the interaction than the other group.

It is worth noting that several participants paid close attention to the technical aspects of the connection, highlighting possible risks that might occur with wearable AR concepts (n=10). Five participants voiced concerns about the potential malfunctions of individual entities, which can imperil the operation of the integrated system. P6, for example, mentioned a scenario where \textit{`one vehicle does not comprehend the signal'}. Three participants suggested connection failures, such as internet disconnections (P11) and signal transmission delays (P14). Two participants highlighted that the system might suffer from malicious manipulation (e.g., hacking).

\textbf{The combined approach provided extra security}. Several participants reported that seeing the zebra crossing and car overlays simultaneously boosted their confidence (n=12). In this regard, they reasoned that the dual cues provided \textit{`extra'} security by exhibiting a strong integration of various entities (i.e., the AR glasses and the vehicles). As P19 explained, \textit{`If there is a misconfiguration or anything that is not synchronized, I may be aware of that and know when the system has an issue'}. 

Further, we noted a number of remarks on the perceived usefulness of each visual cue, shedding light on why their presence was instrumental in the pedestrian crossing experience. Approximately half of the participants interpreted the car overlay as a direct response from each vehicle to their crossing request (n=11). P1, for instance, felt as though \textit{`[the vehicles were] actually listening'} and that the connection worked. In scenarios lacking these individual confirmations, participants reported feeling uncertain about the AV yielding behavior (n=3). As P21 expressed, \textit{`[…] what if there are three or four lanes of cars? If I don't see this green thing, I feel a little bit worried. Maybe some cars will stop, and some will not stop'}. With respect to the zebra crossing, a sizeable proportion of the participants regarded it as a clear crossing signal due to its high visibility (n=3) and familiarity (n=9). The AR marking superimposed on the street also served as a visual cue indicating where the AVs would stop (n=5). 




\section{Discussion}  
In this section, we discuss the findings in relation to our research questions and reflect on the limitations of our study.

\subsection{Preference for Wearable AR Concepts (RQ1)}

The quantitative results indicated that employing wearable AR to aid AV–pedestrian interaction was a viable approach. This is evident in the case of the \textit{AR-combined} concept, which was ranked higher than the baseline pedestrian button and significantly reduced the street-crossing cognitive load. In addition, even though the concept was rated marginally lower in usability and trust due to its unfamiliar nature, no statistically significant differences could be found. However, not all the wearable AR concepts performed similarly. The \textit{AR overlay} and \textit{AR crosswalk} both significantly induced higher distrust and lower trust compared to the baseline; they also received lower usability scores. The discrepancy in the ratings among the wearable AR concepts leads us to infer that the communication approach employed strongly influenced the pedestrians’ subjective experiences. The qualitative feedback confirmed this observation and further suggested that the extent to which pedestrians preferred to use AR glasses to interact with AVs was also influenced by their perception of wearable AR technology. 

With respect to wearable AR technology, the semi-structured interviews revealed important factors influencing pedestrians’ adoption of AR solutions in interacting with AVs, including costs, data privacy, technical defects, liability risks, maintenance duties, and form factors. Although these problems were not widely discussed among the participants, they reinforce expert opinions that wearable AR should not be the sole means for pedestrians to cross the street or engage with AVs in general \citep{tabone2021vulnerable}. Several participants suggested alternative methods of communication with AVs, such as using smartphones, which indicated that a user-initiated communication concept was appreciated more than the underlying AR technology. This inclination might be explained by smartphones’ present ubiquity and their ecosystem of applications. As wearable AR is becoming more pervasive with continuous AR experiences \citep{grubert2016towards} – for example, a pedestrian may use wearable AR for navigational instructions, communication with AVs when crossing the road, or retrieving information about the next train home – we hypothesize that pedestrian attitudes may shift in the future.

Concerning interactions with AVs in safety-critical settings, the participants unanimously agreed on the need to make their crossing intentions known to AVs. This finding is consistent with a prior study on bidirectional communication between pedestrians and AVs \citep{epke2021see, colley2021feedback}, which showed that a combination of hand gestures and receptive eHMIs was the most desired method of communication. However, while hand gestures have been previously observed to have limitations in terms of false-positive \citep{epke2021see} or false-negative detection \citep{gruenefeld2019vroad}, a digital approach was viewed as safer and more trustworthy in our study. Additionally, using wearable AR for bidirectional communication not only ensures that AVs accurately interpret pedestrian intentions but might also eliminate potential confusion about AV non-yielding behaviors \citep{epke2021see}. For example, AR may be utilized to increase system transparency by explaining long wait times or a refusal to yield. According to prior work, explanations of AI system behavior can promote trust in and acceptance of autonomous driving \citep{koo2015did}. A substantial body of literature on explainable AI has focused on drivers’ perspectives; nevertheless, a survey paper has argued that the provision of meaningful explanations from AVs could also benefit other stakeholders (e.g., pedestrians) \citep{omeiza2021explanations}. 

It was anticipated that wearable AR could readily enable targeted and high-resolution communication between AVs and individual pedestrians; a user could be assured that the AVs were addressing them because the device was used individually. In our study, the clarity of recipient was further reinforced when pedestrians were the ones who initiated the communication. However, despite the advantages of wearable AR concepts in delivering unambiguous messages, we found that the aspect of individual perceptions merits further discussion. According to qualitative data, the participants were concerned whether the proposed AR solutions would benefit urban traffic as a whole, as revealed by the raised concerns about frequent crossing requests. In this regard, P17 made a noteworthy comment about a possible shared perception among wearable AR users with regard to visual signals: \textit{`if there are also other people, then I will prefer the crosswalk. People will be crossing the street at the same time and in the same place'}. This comment leads us to believe that in certain situations, a shared AR experience \citep{rekimoto1996transvision}, where multiple users can see the same virtual elements, may help guide pedestrian traffic more efficiently. As a result, personal and shared (augmented) reality should both be considered when designing AR eHMIs.

\subsection{Communication Strategies (RQ2)}
The quantitative findings suggested that the \textit{AR-combined} concept performed better than the \textit{AR crosswalk} (aggregated response) and \textit{AR overlay} (distributed response) across all measures, despite a statistically significant difference only being observed in the concept ranking. In terms of cognitive load, the uncorrected p-values indicated that combining visual signals could considerably reduce pedestrians’ temporal demand as compared to presenting each cue individually, which might mean that when using the \textit{AR-combined} concept, the participants felt less time-pressured as they crossed the road. This tendency was supported by qualitative findings where the participants reported feeling more confident during their crossings. To further understand the benefits and drawbacks of each communication approach, as well as why they were able to complement one another, we discuss them in further detail as follows.

\textbf{(1) Aggregated response}: As one of the most widely recognized traffic symbols, the marked pedestrian crossing was chosen to show an aggregated response from all incoming vehicles, indicating that they were aware of the pedestrians and would yield to them. We expected that this communication strategy would reduce the amount of time and effort required to read implicit or explicit cues from many vehicles. However, the analysis revealed that while the crosswalk indicated a clear signal to cross and a designated crossing area, the participants remained unsure of the AVs’ yielding intention and relied more on vehicle kinematics to make crossing decisions. This finding appears to contradict those of \citet{locken2019should}, in which the participants began crossing as soon as the smart road’s crosswalk lights had turned green, without waiting for AV signals. We believe that the difference in traffic scenarios (one vehicle versus multiple vehicles) and the underlying technologies (smart infrastructure versus personal devices) between the two studies might have contributed to divergent outcomes. 

Notably, user interviews indicated that the participants were not familiar with the notion of connected vehicles; their hesitation persisted even after the leading vehicles had come to a complete stop. The fact that pedestrians do not perceive all AVs as a single system has also been observed during an evaluation of the ``omniscient narrator," where one representative vehicle was in charge of aural communications \citep{colley2020inclusive}. However, it is worth noting that in our study, the crossing signal originated from the AR glasses rather than from one of the AVs. Therefore, it would be useful to further investigate the difference between the two approaches. Moreover, while \citet{colley2020inclusive} expressed reservations about the practicality of aggregated communication in mixed traffic scenarios, we believe that the approach may be feasible with the introduction of connected vehicle technologies. Through the use of in-vehicle or aftermarket devices, vehicles of varying levels of automation can exchange data with other vehicles (V2V), roadside infrastructures (V2I), and networks (V2N) \citep{boban2018connected}. Such connections may result in a gradual shift of pedestrian trust away from specific entities and toward the traffic system as a whole. For example, when responding to the Trust Scale questionnaire, several participants stated that they viewed AR glasses and AVs as a unified system.

\textbf{(2) Distributed response}: The multi-vehicle traffic situation highlighted the necessity for pedestrians to be guaranteed successful communication with every AV, as evident in the positive user feedback on the car overlay. However, a confounding factor was present in the results when some individuals overlooked the overlay, believing that the cars were \textit{`always green'}. We attributed the cause of this issue to the simulation of AR in VR, where the contrast between the augmented graphics and the ``real" environment was not as accurate as it should have been. Additionally, we believe that the participants’ attention might have been scattered in a scenario involving multiple vehicles. For instance, P21 stated that he had to turn left and right to observe the two-way traffic and, therefore, failed to notice \textit{`the changing colors'}. This issue of split attention in complex traffic situations might also present difficulties for distance-dependent eHMIs \citep{dey2020taming}, the encoded states of which change with the distance-to-arrival, as pedestrians may not notice the entire sequence. 

Regarding the display of individual car responses in complex mixed traffic situations, the study findings of \citet{mahadevan2019av} have suggested that this approach would enable pedestrians to assess each vehicle’s awareness and intent and distinguish AVs from other vehicle types. Nonetheless, even standardized eHMI elements could be problematic since each car manufacturer might opt for slightly different designs. As a result, we believe that wearable AR may present a good opportunity for consistent visual communication across vehicles and serve as a clear indicator of their current operation mode (manual versus autonomous) as needed.

\subsection{Limitations and Future Work}
Firstly, the findings of our study drew on the experiences of a small number of university students and young professionals. Although we anticipate comparable outcomes, a larger representative sample would be beneficial, particularly in resolving some borderline quantitative results. Furthermore, past research indicates that cultural differences may cause eHMIs to not have the same favorable effect across countries \citep{weber2019crossing}. Given that the participants in our study largely came from the same cultural background (92\% Vietnamese, 8\% Indian) and had similar habitual traffic behaviors, the feasibility of transferring the wearable AR concepts to differing cultures should be further investigated. Nevertheless, we argue that AR could easily offer personalized experiences, as opposed to vehicle-mounted or infrastructure-based eHMIs.

Secondly, the ecological validity of this work is limited by the use of a VR simulation. The virtual environment could not fully replicate the complex sensory stimuli found in the real world, and the safety associated with VR testing might have influenced individuals to engage in riskier crossing behaviors. Additionally, a few participants expressed anxiety over colliding with physical objects, despite our assurance otherwise. Nonetheless, the majority of the participants responded favorably to the simulation’s realism, stating that they behaved similarly to how they would in the real world. They did not experience any particular motion sickness symptoms and were not affected by the short-distance teleportation implemented at two-thirds of their crossing. Existing literature also suggests that while achieving absolute validity and numerical predictions may not be possible, the VR method can effectively identify differences and patterns \citep{schneider2020virtually}.


Finally, our study employed VR to prototype wearable AR concepts. Although this approach proved useful in overcoming the technical constraints of current AR HMDs, particularly in an outdoor setting, it was challenging for some participants to distinguish superimposed AR graphics from the virtual environment. To some extent, this issue confounded the results of the design concepts with car overlays (the \textit{AR overlay} and the \textit{AR-combined}), possibly causing them to be rated lower than they should have been. However, we believe that it did not invalidate the findings because the order of the four experimental conditions was counterbalanced, and the participants were able to recognize the visual cue in their second encounter. Furthermore, given the possibility of resolving this issue by contrasting display fidelity between AR and VR elements, we recommend that this VR simulation approach be considered in future work. With a large number of proposed AR design concepts in the literature, such as the nine prototypes created by~\citet{tabone2021towards}, comparison studies may provide intriguing insights into how AR systems best facilitate AV–pedestrian interaction.


\section{Conclusion}
This paper has presented novel AR eHMIs designed to assist AV–pedestrian interaction in multi-vehicle traffic scenarios. Through a VR-based experiment, three wearable AR design concepts with differing communication approaches were evaluated against a pedestrian push button baseline. Our results showed that a wearable AR concept highlighting individual AV responses and offering a clear crossing signal is likely to reduce crossing pedestrians’ cognitive load. Further, enabling pedestrians to initiate the communication offered them a strong sense of control. This aspect of user control is currently underexplored in AV external communication research, pointing to important future work in this domain. Finally, the adoption of wearable AR solutions depends on various factors, and it is critical to consider how VRUs without AR devices can interact with AVs safely and intuitively.

\section*{Author Contributions}

TT contributed to Conceptualisation, Data Collection, Data Analysis, Writing (original draft, review and editing). CP and MT contributed to Conceptualisation, Writing (review and editing). YW contributed to Data Analysis, Writing (original draft, review and editing). All authors have read and approved the final manuscript. 

\section*{Funding}
This research is supported by an Australian Government Research Training Program (RTP) Scholarship and through the ARC Discovery Project DP200102604 Trust and Safety in Autonomous Mobility Systems: A Human-centred Approach.

\section*{Acknowledgments}
The authors acknowledge the statistical assistance of Kathrin Schemann of the Sydney Informatics Hub, a Core Research Facility of the University of Sydney. We thank our colleagues Luke Hespanhol and Marius Hoggenmueller for their valuable feedback and all the participants for taking part in this research. 

\bibliographystyle{frontiersinSCNS_ENG_HUMS} 
\bibliography{reference}

\end{document}